\documentclass{article}

\usepackage{arxiv}

\usepackage[utf8]{inputenc} 
\usepackage[T1]{fontenc}    
\usepackage{hyperref}       
\usepackage{url}            
\usepackage{booktabs}       
\usepackage{amsfonts}       
\usepackage{nicefrac}       
\usepackage{microtype}      
\usepackage{lipsum}		
\usepackage{graphicx}
\usepackage{natbib}
\usepackage{doi}
\usepackage{amsmath}

\title{Accelerating wave simulations with neural dispersion correctors}

\author{%
\href{https://orcid.org/0000-0002-6851-2120}{\includegraphics[scale=0.06]{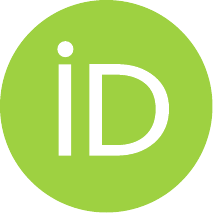}\hspace{1mm}Felipe Rinc\'{o}n}\thanks{Corresponding author: \texttt{felipe.rincon@phd.unipi.it}}\\
Earth Sciences Department\\
University of Pisa\\
56126 Pisa, Italy\\
\texttt{felipe.rincon@phd.unipi.it}
\And
\href{https://orcid.org/0000-0003-3090-963X}{\includegraphics[scale=0.06]{orcid.pdf}\hspace{1mm}Andreas Fichtner}\\
Department of Earth and Planetary Sciences\\
ETH Zurich\\
8092 Zurich, Switzerland\\
\texttt{andreas.fichtner@eaps.ethz.ch}
\And
\href{https://orcid.org/0000-0003-1433-0281}{\includegraphics[scale=0.06]{orcid.pdf}\hspace{1mm}Mattia Aleardi}\\
Earth Sciences Department\\
University of Pisa\\
56126 Pisa, Italy\\
\texttt{mattia.aleardi@unipi.it}
\And
\href{https://orcid.org/0000-0003-3827-8094}{\includegraphics[scale=0.06]{orcid.pdf}\hspace{1mm}Andrea Tognarelli}\\
Earth Sciences Department\\
University of Pisa\\
56126 Pisa, Italy\\
\texttt{andrea.tognarelli@unipi.it}
\And
\href{https://orcid.org/0000-0002-7555-7140}{\includegraphics[scale=0.06]{orcid.pdf}\hspace{1mm}Eusebio Stucchi}\\
Earth Sciences Department\\
University of Pisa\\
56126 Pisa, Italy\\
\texttt{eusebio.stucchi@unipi.it}
}

\renewcommand{\shorttitle}{Neural dispersion correctors}

\hypersetup{
pdftitle={A template for the arxiv style},
pdfsubject={q-bio.NC, q-bio.QM},
pdfauthor={David S.~Hippocampus, Elias D.~Striatum},
pdfkeywords={First keyword, Second keyword, More},
}

\begin{document}
\maketitle

\begin{abstract}
We present a Fourier neural operator network, designed to correct dispersion errors in numerical wave simulations. The neural dispersion corrector enables the replacement of a computationally expensive high-accuracy simulation by a less expensive low-accuracy simulation. In contrast to neural network surrogates that fully replace a wave equation, the neural dispersion corrector has only a weak dependence on the distribution of model parameters, such as wave speeds. Consequently, the network can be trained with a significantly smaller dataset, while still generalising to unseen input parameters. Following a description of the network architecture and training, we provide examples for the 3-D elastic wave equation. After training with merely 1$\,$000 examples on one GPU, the neural corrector achieves a speed-up of 16$\times$ compared to a reference spectral-element simulation and a generalisation to a broad range of strongly heterogeneous wave speed distributions.
\end{abstract}


\section{Introduction}
Simulations of wave propagation are the central element of numerous applications in acoustics, medical ultrasound, non-destructive testing, tsunami warning, seismic imaging, and many other domains. Reflecting this importance, a wide range of numerical methods has been developed to approximate solutions to various wave equations for media with arbitrary heterogeneities or complicated geometries, for which closed-form solutions are unavailable. Often striking a good balance between accuracy, versatility and ease of implementation, grid- or element-based methods using finite-difference \citep[e.g.,][]{Bohlen_2002,Moczo_2014}, spectral-element \citep[e.g.,][]{Faccioli_1996,Komatitsch_1998} or discontinuous-Galerkin \citep[e.g.,][]{Dumbser_2007,DeLaPuente_2007} discretisations, are among the most widely used simulation tools.\\[5pt]
Despite their success in modeling a broad spectrum of wave phenomena, these methods share a poor scaling with the highest frequency, $f_\text{max}$, that one would like to simulate. To achieve useful accuracy, a certain number of grid points per minimum wavelength is required. Consequently, the total number of grid points scales as $f_\text{max}^d$, where $d$ is the spatial dimension. The Courant-Friedrichs-Levy (CFL) condition \citep[e.g.,][]{Quarteroni_2000,Igel_2016} furthermore imposes that the length of a time step, $\Delta t$, must decrease linearly with the grid spacing, $\Delta x$. These two effects combine into an unfavourable frequency scaling of $f_\text{max}^{d+1}$. Hence, in 3-D, a doubling of the maximum frequency increases computational cost by a factor of at least $\sim$16. Although computational resources have increased substantially in recent years, many wave simulations cannot be performed at the frequencies where we have data or would like to compute predictions.\\[5pt]
In tandem with the development of deep learning for scientific problem solving, several approaches have been proposed to reduce the computational cost of numerical wave simulations by replacing wave equations by neural networks with different architectures. Examples include physics-informed neural networks (PINNs) and Fourier neural operators (FNOs). PINNs complement the loss function with a term that forces the network to solve the PDE of interest \citep{Raissi_2019}, which can improve generalisation. Fourier neural operators (FNOs) are designed to learn the underlying PDE, which makes them grid- and scale-independent \citep{kovachki_2021}. Current adaptations of PINNs and FNOs to wave equations employ the acoustic approximation \citep[e.g.,][]{Moseley_2020,Rasht_2022}, are limited to 2-D \citep[e.g.,][]{Moseley_2020,Rasht_2022,Ren_2024} or propagate only few wavelengths in 3-D \citep[e.g.,][]{Song_2022,Zou_2024}.\\[5pt]
A limitation that PINNs and FNOs have in common is the required size of the training dataset. Although there is no precise and universal scaling relation, the required number of training samples is expected to scale with the effective number of model parameters to which the wavefield is sensitive or which it can resolve. The diffraction limit imposes that a wave with wavelength $\lambda$ may resolve structure as small as $\lambda/2$. Consequently, the effective size of the model parameter space likely grows as $1/\lambda^d$ or $f_\text{max}^d$. Although PINNs compensate effective parameter space growth to some extent by forcing the network to solve the PDE, this comes with the potentially prohibitive cost of evaluating numerous second derivatives of the network (or a larger equivalent system of first order) during network training. As a result, neither PINNs nor FNOs currently operate at the scales of recent massive wave simulation and inversion problems that propagate waves over hundreds or thousands of wavelengths in applications ranging from seismic imaging to medical ultrasound \citep[e.g.,][]{Angla_2023,Thrastarson_2024,Cui_2024,Marty_2024,Metivier_2025}.\\[5pt]
The poor scaling of the training dataset for standard PINN or FNO surrogates motivates the search for an alternative approach that is less dependent on the size of the effective model parameter space. In this context, we propose a hybrid approach that combines grid- or element-based numerical solutions with neural networks. The basic concept is to employ only a small dataset for training a neural network capable of correcting numerical dispersion errors incurred by running computationally less expensive numerical simulations with an overly coarse grid. If successful, the method would allow us, for example, to perform sloppy 3-D simulations with twice the grid spacing of the accurate simulation with a $1/16$ times the computational cost, and subsequently repair the dispersion error to retrieve accurate results. Our contribution extends the work of \citet{Agnihotri_2022} on 2-D acoustic neural correctors by adding the ability to simulate 3-D vectorial (e.g., elastic) waves, a theoretical justification for training efficiency, and detailed performance analysis.\\[5pt]
This manuscript is structured as follows: In section \ref{S:Dispersion} we demonstrate that dispersion errors depend only weakly on the medium properties, thereby providing a theoretical justification for the small size of the dataset needed to train the neural dispersion corrector. The design of the corrector is the topic of section \ref{S:Correctors}, where we introduce our example problem (the 3-D elastic wave equation) and describe the network architecture and parsimonious data representation. Section \ref{S:Results} provides concrete examples of training data preparation, network training, validation and generalisation. This is complemented by section \ref{S:Limitations} on limitations related to incorrect medium parameters and simulations beyond the training dataset. Finally, section \ref{S:Discussion} discusses the balance between training and simulation cost, numerical accuracy, and the relation to analytical correctors of temporal discretisation correctors.

\section{Dispersion errors and their dependence on medium properties}
\label{S:Dispersion}

\begin{figure}
	\centering
	\includegraphics[width=0.9\columnwidth]{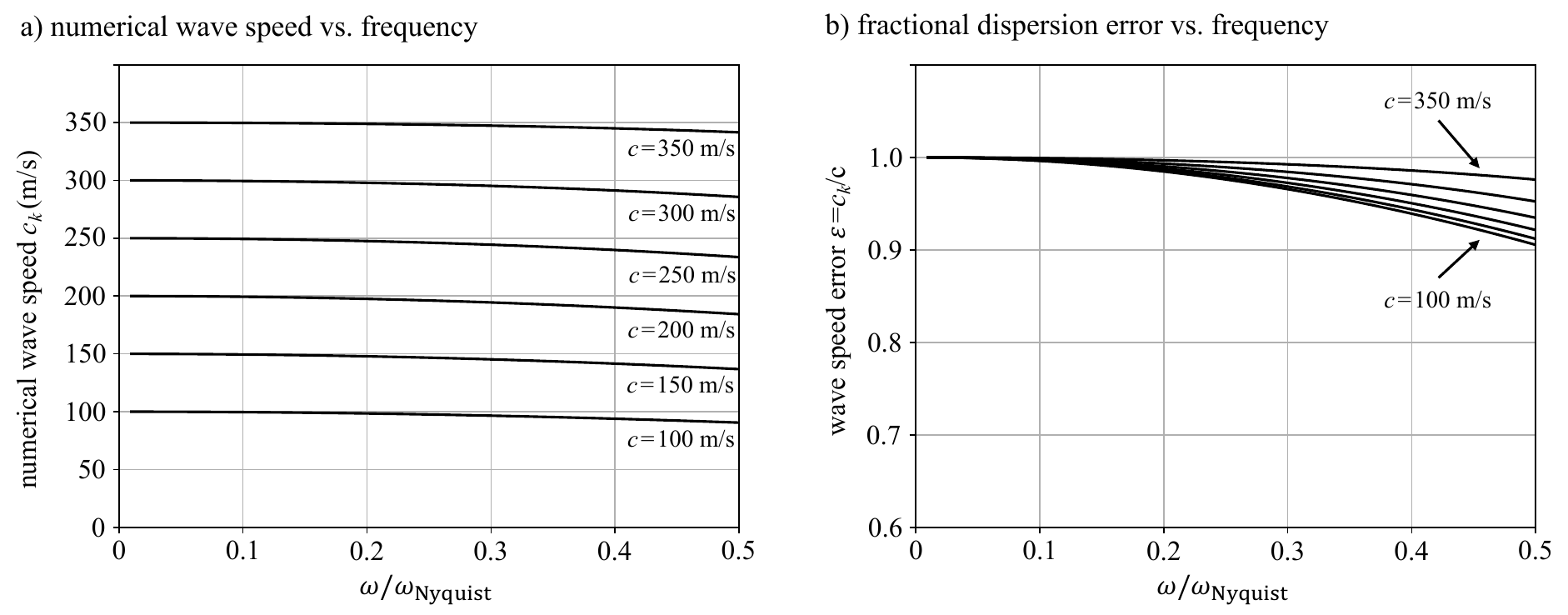}
	\caption{Comparison of numerical wave speed $c_k$ (a) and fractional wave speed error $c_k/c$ (b) as a function of frequency, normalised with respect to the Nyquist frequency $\omega_\text{Nyquist}$. The discretisation is fixed to $dx$=100 m and $dt$=0.25. Different curves correspond to different exact wave speeds $c$. In contrast to $c_k$ itself, the fractional error only shows a weak dependence on $c$, despite its large variation from 100 to 350 m$/$s. The dependence is particularly weak in the regime where most second-order finite-difference simulations operate, i.e., at $\omega/\omega_{Nyquist}\leq 0.2$, which corresponds to $\geq$10 grid points per wavelength \citep[e.g.,][]{Moczo_2014,Igel_2016}.}
	\label{fig:error}
\end{figure}

The centrepiece of our approach is the training of a dispersion correction network with only a small training dataset that can be computed with a cost that is lower than or comparable to the cost of a typical full-waveform inversion. The attainability of this goal can be justified with a standard von Neumann analysis, which reveals that the numerical dispersion error is nearly independent of the wave speed distribution. Since numerous text books treat von Neumann analysis extensively \citep[e.g.,][]{Quarteroni_2000,Fichtner_book,Igel_2016}, we restrict ourselves to a condensed line of arguments, for which we consider the simplest case of a 1-D wave equation
\begin{equation}\label{E:neumann001}
\frac{\partial^2 u(x,t)}{\partial t^2} - c^2\,\frac{\partial^2 u(x,t)}{\partial x^2} = 0\,,
\end{equation}
where $u(x,t)$ is the wavefield as a function of space $x$ and time $t$, and $c$ is the wave speed. We can express the approximate solution at a discrete set of spatial grid points $x_n = n\,\Delta x$, spaced at equal distance $\Delta x$ and indexed by the integer $n$, in the form of a Fourier series,
\begin{equation}\label{E:neumann002}
u_n(t) = \sum_k  \psi_k(t) \, e^{i\,k n\,\Delta x}\,,
\end{equation}
with the time-dependent Fourier coefficients $\psi_k(t)$. Although the sum over wave numbers $k$ runs from $-\infty$ to $\infty$, temporal frequencies corresponding to wave numbers beyond the Nyquist limit $k_\text{max}=\pi/\Delta x$, are typically eliminated from the numerical solution by low-pass filtering. Replacing the second time and space derivatives with second-order central finite differences and inserting $u_n$, results in an explicit equation for the Fourier coefficients at the $j^\text{th}$ time step $t_j = j\Delta t$\,,
\begin{equation}\label{E:neumann003}
\psi_k(j\Delta t) = \psi_0 e^{i\, k n \,\Delta x}\, \lambda_k^j\,.
\end{equation}
The complex-valued numbers $\lambda_k$ are given by
\begin{equation}\label{E:neumann004}
\lambda_k = \frac{1}{2} \left(a_k \pm \sqrt{a_k^2 -4}\right)\,,
\end{equation} 
with 
\begin{equation}\label{E:neumann005}
a_k = 2 - 4 c^2 \frac{\Delta t^2}{\Delta x^2} \sin^2\left( \frac{1}{2} \frac{\omega_k}{c}\, \Delta x  \right)\,,
\end{equation}
and the angular frequency $\omega_k = k c$. When the time step $\Delta t $ meets the CFL condition $\Delta t < \Delta x / c$, we obtain $|\lambda_k|=1$, the simulation becomes numerically stable, and we can write $\lambda_k$ in the form
\begin{equation}\label{E:neumann006}
\lambda_k = e^{i\,\omega_k\Delta t\,c_k / c}\,,
\end{equation}
with the phase of $\lambda_k$ denoted by $\varphi_k$. Equation (\ref{E:neumann006}) implicitly defines the numerical wave speed $c_k = c \varphi_k /(\omega_k \Delta t)$ at frequency $\omega_k$ in terms of the fixed exact wave speed $c$ and the time step $\Delta t$. Our interest is in the fractional wave speed dispersion error 
\begin{equation}\label{E:neumann007}
\varepsilon_k = \frac{c_k}{c} = \frac{\varphi_k}{\omega_k \Delta t}\,, 
\end{equation}
and specifically on its dependence on $c$ itself. To explore this dependence, we expand $a_k$ from (\ref{E:neumann007}) into a Taylor series,
\begin{equation}\label{E:neumann008}
a_k = 2 - \omega_k^2 \Delta t^2 + \frac{1}{12} \frac{\omega_k^4}{c^2} \Delta x^2 \Delta t^2 + \mathcal{O}\left( \frac{\omega_k^6}{c^4} \Delta x^4 \Delta t^2 \right)\,.
\end{equation}
Equation (\ref{E:neumann008}) reveals that $a_k$ has no leading-order dependence on $c$ for a fixed discretisation in terms of $\Delta t$ and $\Delta x$, and a fixed angular frequency $\omega_k$. A perturbation $\delta c$ translates to a perturbation $\delta a_k \propto \delta c/c^3$, meaning that it is small even for large fractional perturbations $\delta c/c$. Furthermore, the dependence on $c$ can be reduced quickly by decreasing $\omega_k$. Via the variation of (\ref{E:neumann005}), these properties translates to $\lambda_k$, its phase $\varphi_k$, and the fractional error $\varepsilon_k$, i.e.,
\begin{equation}\label{E:neumann009}
\delta\varepsilon_k \propto \frac{\delta c}{c^3}\,.
\end{equation}
Hence, numerical dispersion is primarily a property of the grid and not a property of the medium through which the wave propagates. Fig. \ref{fig:error} visualises the dependence of $\varepsilon_k$ on $c$ without any Taylor approximation. It confirms that the dependence is weak, especially in range where second-order finite-difference methods typically operate to achieve highly accurate results, that is, with around 10 grid points per minimum wavelength or more.\\[5pt]
The previous analysis is simplified in that it uses a 1-D wave equation with constant wave speed and one of the most basic spatio-temporal discretisation schemes. Nevertheless, the weak dependence of the dispersion error $\varepsilon_k$ on the wave speed $c$ is in stark contrast to the wavefield $u$ itself, which has a strong dependence on $c$. This suggests that the dispersion error may be a more suitable objective for neural network training than the full wavefield. 
%

\section{Neural dispersion correctors}\label{S:Correctors}

Based on the analysis in section \ref{S:Dispersion}, we aim to design a neural network that can be trained with a small dataset, consisting of low-accuracy wavefields as input and high-accuracy wavefields as output. Our network design, described in detail in section \ref{SS:Design}, repurposes FNOs from operator surrogates into upscaling tools. We train the FNO to map a low-accuracy wavefield -- affected by numerical dispersion errors due to coarse discretisation -- into its high-accuracy counterpart, which would otherwise require a more expensive simulation. This design choice reflects a pragmatic balance. Rather than relearning the complete underlying physics from data, we exploit the efficiency and resolution-invariance of the FNO to correct the errors introduced by coarse numerical modeling. By focusing on the low-accuracy $\rightarrow$ high-accuracy mapping, the network effectively acts as a dispersion error corrector, delivering wavefields that retain the fidelity of fine-scale simulations at the cost of coarse ones. In this section, we describe the details of the network architecture, the compact dataset representation adopted for training, and the metrics used to evaluate the upscaling performance.

\subsection{Example problem}\label{SS:Example}

Before we delve into the details of network design, we briefly summarise the example problem that we will use for illustration in later sections. To ensure that our approach can be applied to a wide range of wave equations by reducing the complexity of the network instead of having to increase it, we consider the 3-D elastic wave equation
\begin{equation}\label{E:wave000}
\rho(x) \frac{\partial^2 u_i(x,t)}{\partial t^2} - \sum_{j,k,l=1}^3 \frac{\partial}{\partial x_j} \left[ C_{ijkl}(x) \frac{\partial}{\partial x_k} u_l(x,t)  \right] = \delta(x-x_0)\,s(t)\,,
\end{equation}
where $u_{i=1,2,3}$ are the components of the wavefield $u \in \mathbb{R}^3$ that depend on space $x \in G \subset \mathbb{R}^3$ and time $t \in [0, T]$. The right-hand side of (\ref{E:wave000}) specifies the properties of a point-localised source in terms of its position $x_0$ and source wavelet $s(t)$. The material parameters are mass density $\rho(x)$ and the elastic tensor $C_{ijkl}(x)$ \citep[e.g.,][]{Kennett_2001,Aki_Richards_2002}. In an isotropic medium, $C_{ijkl}$ takes the special form $C_{ijkl}=\lambda \delta_{ij}\delta_{kl}+\mu\delta_{ik}\delta_{jl}+\mu\delta_{il}\delta_{jk}$, with the Lam\'{e} parameters $\lambda$ and $\mu$. Along the physical boundary of the domain $G$, we complement (\ref{E:wave000}) by free-surface conditions. Absorbing boundary conditions are implemented along the unphysical part of the boundary, which results from truncating the domain to a computationally tractable size. Solutions of (\ref{E:wave000}) are vector-valued waves, including Rayleigh and Love surface waves, as well as longitudinal body waves (P waves) and transverse body waves (S waves) with propagation speeds $v_\text{p} = \sqrt{(\lambda + 2\mu)/\rho}$ and $v_\text{s}=\sqrt{\mu/\rho}$, respectively \citep{LandauLifshitz_1986}. In the examples in section \ref{S:Results} and \ref{S:Limitations}, we employ the spectral-element code Salvus \citep{Afanasiev_2019} to solve (\ref{E:wave000}) numerically for arbitrary distributions of $\rho$, $v_\text{p}$ and $v_\text{s}$.\\[5pt] 
Equation (\ref{E:wave000}) is more complex than several other widely used wave equations. Isotropic electromagnetic wave propagation is described by a simpler variant of (\ref{E:wave000}), where the elastic tensor is replaced by a single scalar. The acoustic wave equation is a special case that emerges from (\ref{E:wave000}) by setting $\mu=0$ and solving only for the scalar pressure field $p=-\nabla\cdot u$. Continuing from the acoustic wave equation, the Schr\"{o}dinger equation can be obtained via different linear transformations \citep[e.g.,][]{Babbush_2023,Schade_2024}.

\subsection{Network design}\label{SS:Design}

\begin{figure*}
	\centering
	\includegraphics[width=\columnwidth]{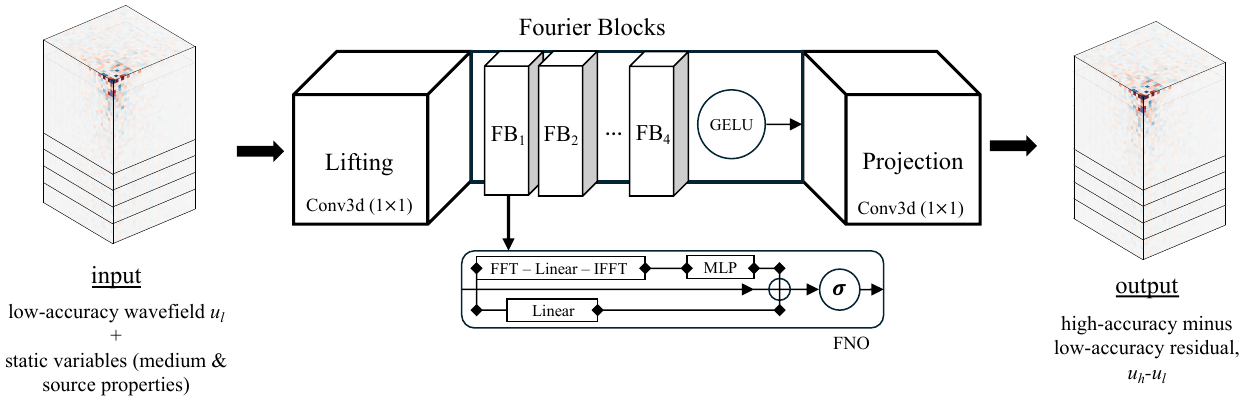}
	\caption{Schematic illustration of the FNO architecture used in the following numerical examples. The input consists of low-accuracy three-components wavefields in the frequency domain and static variables that describe elastic medium and source properties. These are mapped into a higher-dimensional representation through a lifting layer. Each FB consists of Fourier transforms followed by a pointwise multi-layer perceptron (MPL). After passing through four FBs, the final projection layer maps the latent representation back into six output channels, corresponding to the real and imaginary parts of the three-component frequency-domain residual between the high-accuracy wavefield $u_h$ and its low-accuracy counterpart $u_l$.}
	\label{fig:fno_architecture}
\end{figure*}

Neural operators (NOs) are a class of machine learning models designed to approximate the mapping of some operator from an input to an output function space \citep{lu_2021, kovachki_2023}. Unlike standard neural networks that approximate finite-dimensional functions, NOs target infinite-dimensional mappings. Despite being trained on a finite amount of discrete data, NOs act on functions and are therefore resolution- and mesh-invariant \citep{kovachki_2021}. Once trained, NOs can take input from an arbitrarily discretised function and produce consistent output, thereby enabling tasks such as the transfer of solutions between meshes.\\[5pt]
An FNO is a particular NO that parameterises an integral kernel in Fourier space \citep{li_2020}. Each network layer transforms the input field via a Fourier transform, applies a learned linear map on a truncated set of Fourier modes, and then transforms back. Combined with local pointwise operations and nonlinearities, this design captures both long-range interactions through spectral multipliers and local nonlinear structure through activations. The result is an architecture that is conceptually simple, fast to implement using fast Fourier transforms (FFTs), and effective in learning mappings between infinite-dimensional function spaces. In practice, FNOs have shown high accuracy across PDE benchmarks, while being orders of magnitude faster than conventional numerical solvers \citep{li_2022}.\\[5pt]
Our FNO network $\mathcal{N}$ learns to correct the dispersion error in a low-accuracy simulation $u_l$. The network output $\mathcal{N}u_l$ upscales $u_l$ such that it approximates the high-accuracy version $u_h$. To obtain a more compact representation of the network in- and output, we transform the time-dependent wavefield into the frequency domain and select a discrete set of frequencies, similar to the approach of \citet{Zou_2024}. Hence, the network has six input channels for the wavefield, consisting of the real and imaginary parts of the spectrum at a certain frequency for each of the three components. Furthermore, the network receives a set of static variables, consisting of the medium properties $v_\text{p}$, $v_\text{s}$ and $\rho$, a 3-D Gaussian function with its maximum at the source location, and the central frequency of the source wavelet used for the simulations. Instead of directly producing an upscaled wavefield, the output of the network consists of the real and imaginary parts of the three components of the corrector residual $\Delta u= u_h - u_l$. \\[5pt]
Our network architecture can be decomposed into three parts, summarised in Fig.~\ref{fig:fno_architecture}:  
(i) a lifting layer that projects the input channels into a higher-dimensional latent space;  
(ii) a sequence of Fourier blocks (FBs) that perform truncated 3-D Fourier convolutions combined with nonlinear activation; and  
(iii) a projection layer that maps the latent representation back to the output channels. Denoting by $h_\ell$ the latent features at the input of the $\ell^\text{th}$ FB, we define the spectral operator of this block as
\begin{equation}\label{E:fno000}
\mathcal{S}_\ell(h_\ell) \;=\; \mathcal{F}^{-1}\!\left[\,T_\ell \Big( R_\ell \,\mathcal{F}\{h_\ell\}\Big)\right],
\end{equation}
where $\mathcal{F}$ and $\mathcal{F}^{-1}$ denote the 3-D FFT and its inverse, $R_\ell$ are the learnable complex-valued weights acting in the spectral domain, and $T_\ell$ is a truncation operator that retains only a fixed number of low-frequency modes.  
The FB update is then given by
\begin{equation}\label{E:fno001}
h_{\ell+1} \;=\; \sigma\!\Big(\mathcal{S}_\ell(h_\ell) \;+\; W_\ell h_\ell \;+\; b_\ell\Big),
\end{equation}
where $W_\ell h_\ell + b_\ell$ denotes a learnable convolution applied in the spatial domain, and $\sigma(\cdot)$ is the nonlinear activation function. Following empirical tests, we use the GELU activation \citep[e.g.,][]{Hendrycks_2023} as activation function that empirically performs well. Fig.~\ref{fig:fno_architecture} provides a schematic overview of the network architecture. The lifting layer projects the low-resolution wavefield $u_l$ together with the static variables into a latent space of 48 channels. Each FB then performs a truncated spectral convolution that retains 16 modes in each dimension, adds a parallel spatial convolution, and applies a GELU activation. Finally, the projection layer maps the 48 latent channels back to 6 output channels, which represent the residual $\Delta u$ for the real and imaginary parts of the three wavefield components. As described in section \ref{S:Parsimonious}, we implement the option of compressing the frequency-domain wavefields using a discrete cosine transform (DCT) in order to reduce storage and memory requirements.
%

\subsection{Parsimonious data representation}\label{S:Parsimonious}

Our proposed neural corrector is designed to mitigate the numerical dispersion errors by learning a mapping from computationally inexpensive, low-accuracy simulations to their high-accuracy counterparts. The intended key advantage is that training can be performed with relatively small datasets while still achieving strong generalisation across a wide range of scenarios. Nevertheless, each training example requires both a low-accuracy and a high-accuracy simulation, which may introduce a bottleneck in terms of computational cost and memory usage.\\[5pt] 
Owing to the CFL condition, the total number of space–time grid points in a wavefield simulation scales as $N^{d+1}$, where $N$ is the number of grid points per spatial dimension and $d$ is the number of spatial dimensions. In special cases where the medium is smooth and nearly rotationally symmetric, wavefield-adapted meshes can improve this scaling \citep[e.g.,][]{vanDriel_2020,Thrastarson_2020}. However, such dimensionality reduction strategies tend to fail in important applications such as surface- and guided-wave modeling or non-destructive testing, where strong medium contrasts and small-scale heterogeneities are common. Hence, memory bottlenecks and time-stepping constraints often remain major obstacles in network training. In addition to the cost of computing the high-accuracy reference data, the RAM and GPU memory required to store and process full-resolution wavefields during training can quickly become prohibitive \citep{kovachki_2023}.\\[5pt] 
To maintain a fast and memory-efficient framework, we combine two strategies. First we perform the simulations in the time domain and subsequently transform the wavefields into the frequency domain. By training directly on discrete frequencies, the neural corrector effectively treats each frequency independently, thereby avoiding explicit time discretization (\citet{Zou_2024}). Second, we compress both real and imaginary components of the frequency domain wavefields using the discrete cosine transform. This compact representation substantially reduces memory and training costs, while retaining the essential physical information. We adopt the DCT for its conceptual simplicity, linearity, and efficient implementation in both forward and inverse form. Recent studies have demonstrated the potential of incorporating the DCT in neural network training for geophysical problems \citep[e.g.,][]{Aleardi_2021, Rincon_2025a, Rincon_2025b}. In section \ref{S:Accuracy}, we provide a more detailed discussion on the tuning of DCT compression.

\section{Results}\label{S:Results}

To demonstrate the applicability of our neural corrector, we present results for near-surface seismic wave propagation using the 3-D elastic wave equation (\ref{E:wave000}) in a domain with dimensions of $100 \times 100 \times 40 \,\mathrm{m}^3$. Such cases are particularly challenging because low wave speeds combined with strong wave speed contrasts require high-accuracy simulations, which could lead to excessive memory and computational time requirements for training. This experiment specifically highlights the practicality of the reduced dataset representation, discussed in section \ref{S:Parsimonious}.

\subsection{Training dataset}\label{S:Data}

\begin{figure}
	\centering
	\includegraphics[width=1.0\columnwidth]{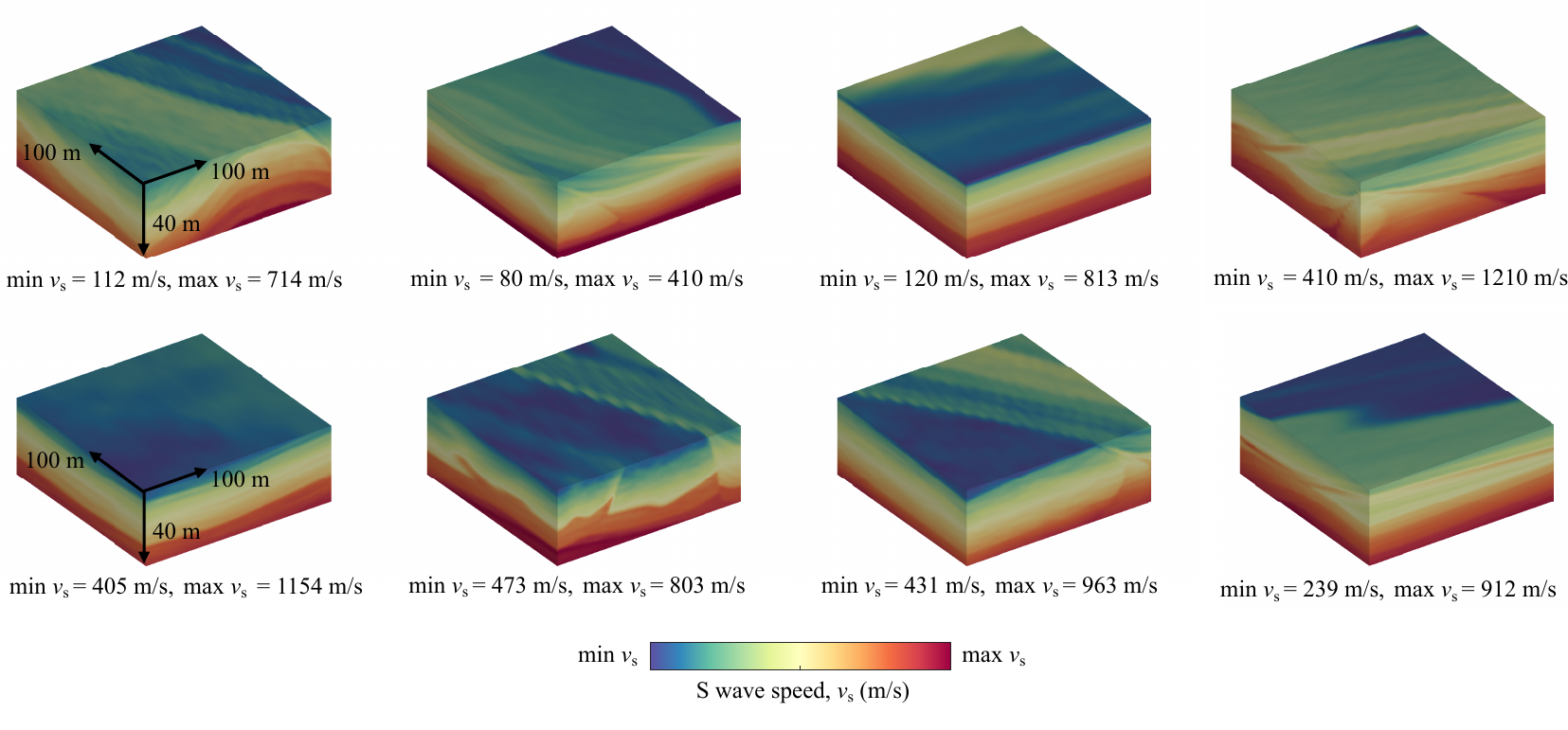}
	\caption{Representative selection of eight of the 100 $v_\text{s}$ distributions that we used to compute $v_\text{p}$ and $\rho$, as well as the high- and low-accuracy wavefields that compose the training dataset.}
	\label{fig:dataset}
\end{figure}

\begin{figure}
	\centering
	\includegraphics[width=0.9\columnwidth]{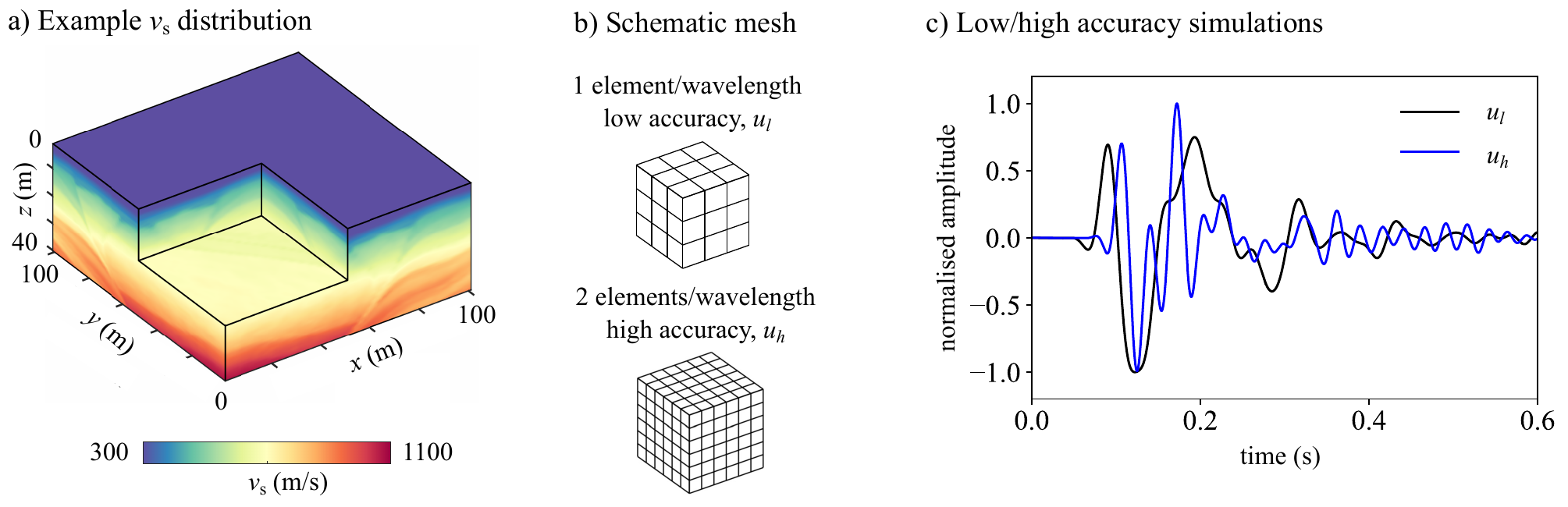}
	\caption{Comparison of low- and high-accuracy spectral-element simulations. Panel (a) shows the near-surface $v_\text{s}$ distribution used for this example. Panel~(b) illustrates a close-up of the discretisation. Doubling the elements per wavelength (epw) from one to two, the number of elements increases from 1$\,$859 to 13$\,$125. Finally, panel~(c) compares representative vertical-component time series from the low-accuracy wavefield $u_l$ (black) and the high-accuracy wavefield $u_h$ (blue). The goal of the neural dispersion corrector is to map $u_l$ into a close approximation of $u_h$.}
	\label{fig:mod_sim}
\end{figure}

To build a suitable training dataset, we first create a set of 100 $v_\text{s}$ distributions that are representative of typical near-surface structures. These structures, a selection of which is shown in Fig.~\ref{fig:dataset}, include geologically plausible wave speeds in the range of 100 to 1200~m~s$^{-1}$. For each $v_\text{s}$, we derive $v_\text{p}$ and $\rho$ using common empirical scaling relations \citep{Brocher_2005, Castagna_1985, Gardner_1974}, as well as the theoretical relations between density and wave speeds, mentioned in section \ref{SS:Example}.\\[5pt]  
%
%
Implementing a free surface at the top and absorbing boundaries at the lateral and bottom faces, we compute three-component time-domain wavefields using the spectral-element solver Salvus \citep{Afanasiev_2019}. The Lagrange polynomial degree in all elements is set to 2, and the total duration of the simulations is T=1 s, which ensures that the wavefield traverses the entire domain. We place moment tensor point sources at the surface and employ three different source wavelets (Ricker, Morlet, and Ormsby) with variable central frequencies of 5, 10, 15, and 20~Hz. By randomly selecting source positions, wavelets and central frequencies, we perform a total of 1$\,$200 simulations in high-accuracy mode with two elements per minimum wavelength. This is complemented by 1$\,$200 simulations of lower accuracy, with only one element per minimum wavelength.\\[5pt]
To obtain a compressed version of the network input, we first compute the FFT of the time-domain wavefield. Instead of using all available frequencies, we select 64 equally spaced frequencies around the central frequency of the source wavelet, and store the real and imaginary parts separately. This frequency-domain subsampling reduces the storage requirement for the wavefield by 97.9 \%. Subsequently, we apply a DCT to both the frequency-domain wavefields and the material parameter distributions. Retaining only the lowest 30 coefficients per component, acts as a spatial low-pass filter with a cutoff close to the minimum wavelength. We provide a more detailed discussion of this choice in section \ref{S:Accuracy}. As a result of DCT filtering, storage and memory requirements decrease by 87.1 \%. Frequency subsampling and DCT filtering combined, compress the wavefield by 99.7 \%.\\[5pt] 
Of the 1$\,$200 compressed wavefields, we use 80 \% for training, 10 \% for validation and 10\% for testing. Finally, we ensure that the testing set only includes parameter distributions and source locations that were not used for training or validation.\\[5pt]
A comparison of the low- and high-accuracy simulations for a model with minimum $v_\text{s}$=200 m/s illustrates the computational savings that the neural corrector could potentially bring. For a source with central frequency of 10~Hz, the low-accuracy mesh contains $\sim$1$\,$900 elements, compared to $\sim$13$\,$100 elements in the high-accuracy version. With the near perfect scaling of Salvus, the corresponding ratio of consumed node hours is $\sim$16. Fig. ~\ref{fig:mod_sim} summarises this setup and its implications for solution accuracy. 
%

\subsection{Learning performance and computational gains}\label{sec:train_val}

\begin{figure}
  \centering
  \includegraphics[width=0.8\columnwidth]{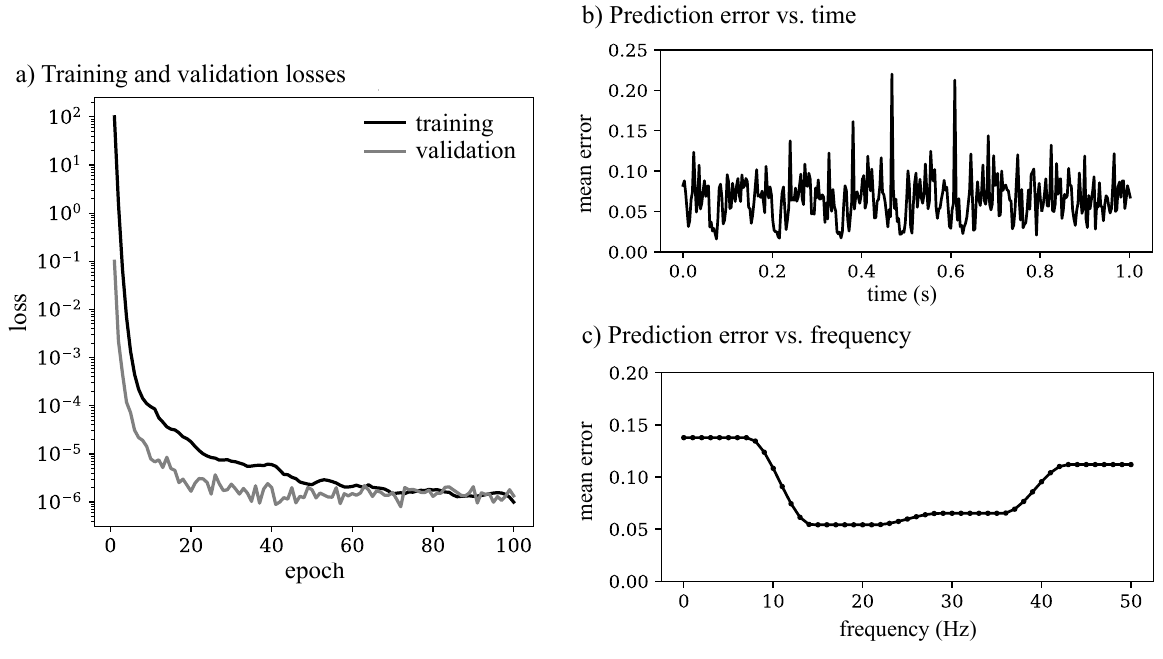}
  \caption{Summary of training and prediction performance. (a) Training and validation losses as a function of Adam optimisation epoch. (b) Mean prediction error for the testing dataset as a function of time, and (c) as a function of frequency.}
  \label{fig:train_val_testing}
\end{figure}

We trained the neural dispersion corrector for 100 epochs with a batch size of 4, a learning rate of $3\times 10^{-4}$ (including one warm-up epoch), and the Adam optimiser \citep{Kingma_2014} with weight decay of $10^{-5}$. The network features 48 channels, 4 Fourier blocks and 16 truncated Fourier modes per spatial dimension. Fig. ~\ref{fig:train_val_testing}a illustrates the evolution of training and validation losses. Both curves decrease smoothly and plateau after $\sim$50 epochs, indicating stable convergence without signs of overfitting. The training took 4 hours and 28 minutes on a single NVIDIA GeForce RTX 3070 Ti GPU.\\[5pt]
Here we quantify the training performance using an error metric defined as one minus the structural similarity index, $1-\mathrm{SSIM}$ \citep{Wang2004SSIM}. Fig. ~\ref{fig:train_val_testing}b summarises the mean prediction error over time for the testing dataset, i.e., the simulations that entered neither the training nor the validation. It remains low and largely steady throughout the wavefield evolution, with only occasional spikes. The mean error for the testing dataset as a function of frequency is shown in Fig.~\ref{fig:train_val_testing}c. The neural corrector achieves its highest accuracy at mid frequencies between 10 - 40 Hz, and is modestly less accurate at the lowest and highest frequencies. This behaviour is consistent with the spectral truncation in the Fourier layers, which emphasises energy-dominant modes.\\[5pt]
Once trained, the network corrects the dispersion error of a low-accuracy simulation in $\sim$1 \% of the time needed to run the simulation, meaning that the computational overhead of the neural corrector is negligible. Given the nearly perfect scaling of Salvus \citep{Afanasiev_2019}, this results in a total reduction of computational requirements by a factor of $\sim$$2^4 = 16$, while introducing only negligible error. In the following section, we present the results of the corrected simulations and discuss the prediction accuracy in more detail.\\[5pt]
%


\subsection{Wavefield prediction: The neural corrector in action}\label{sec:wf_corrections}

\begin{figure}
  \centering
  \includegraphics[width=1.0\columnwidth]{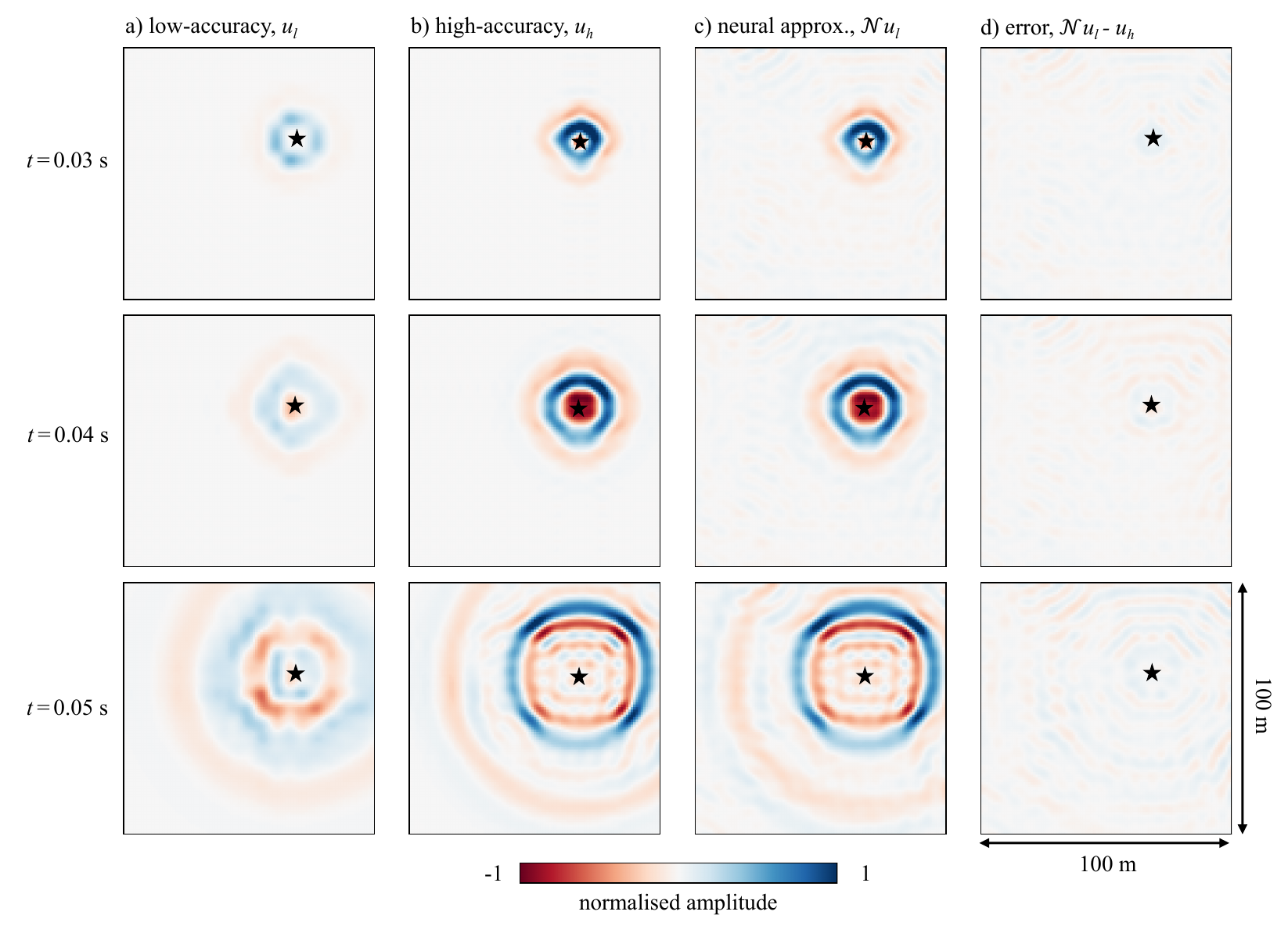}
  \caption{Time snapshots of the vertical wavefield component at the surface. The black star marks the source location. Amplitudes are normalised to the maximum amplitude over all grid points, all times and wavefield examples. (a) Low-accuracy/low-cost wavefield $u_l$ with one second-order element per wavelength. (b) High-accuracy/high-cost reference wavefield $u_h$ with two second-order elements per wavelength. (c) Application of the neural corrector to the low-accuracy wavefield, $\mathcal{N}u_l$. (d) Approximation error $\mathcal{N}u_l - u_r$. The neural correction substantially reduces dispersion errors and closely approximates the high-resolution solution.}
  \label{fig:wavefields}
\end{figure}

\begin{figure}
  \centering
  \includegraphics[width=0.9\columnwidth]{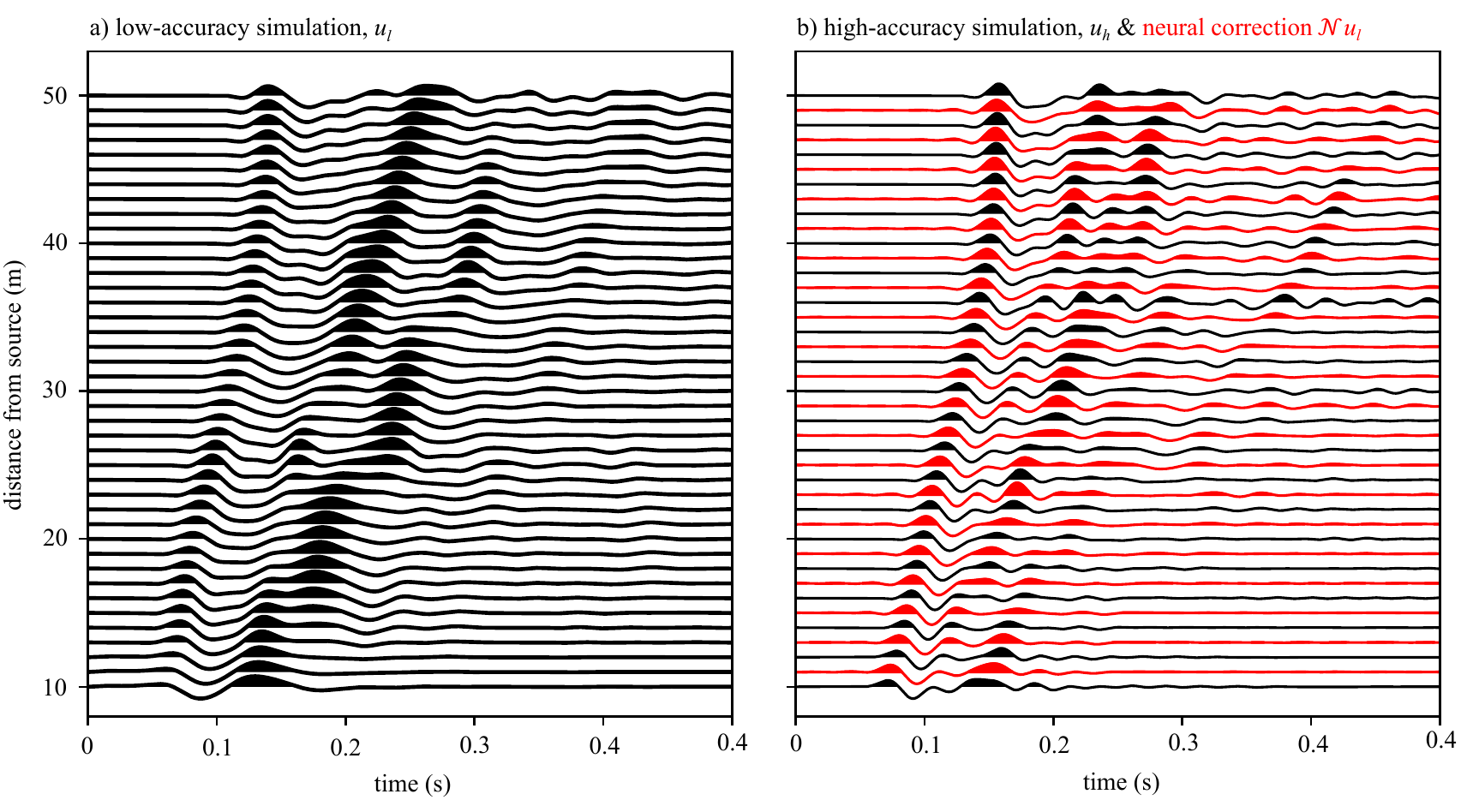}
  \caption{Trace-by-trace comparison of vertical-component wavefields at receivers located between 10 and 50 m distance from the source. a) Low-accuracy/low-cost simulation, $u_l$. The overall frequency content is too low; the traces lack detail. b) Alternating plot of the high-accuracy/high-cost reference $u_h$ (black) and the neural correction $\mathcal{N}u_l$ (red).}
  \label{fig:seismogramLAHA}
\end{figure}

\begin{figure}
  \centering
  \includegraphics[width=0.8\columnwidth]{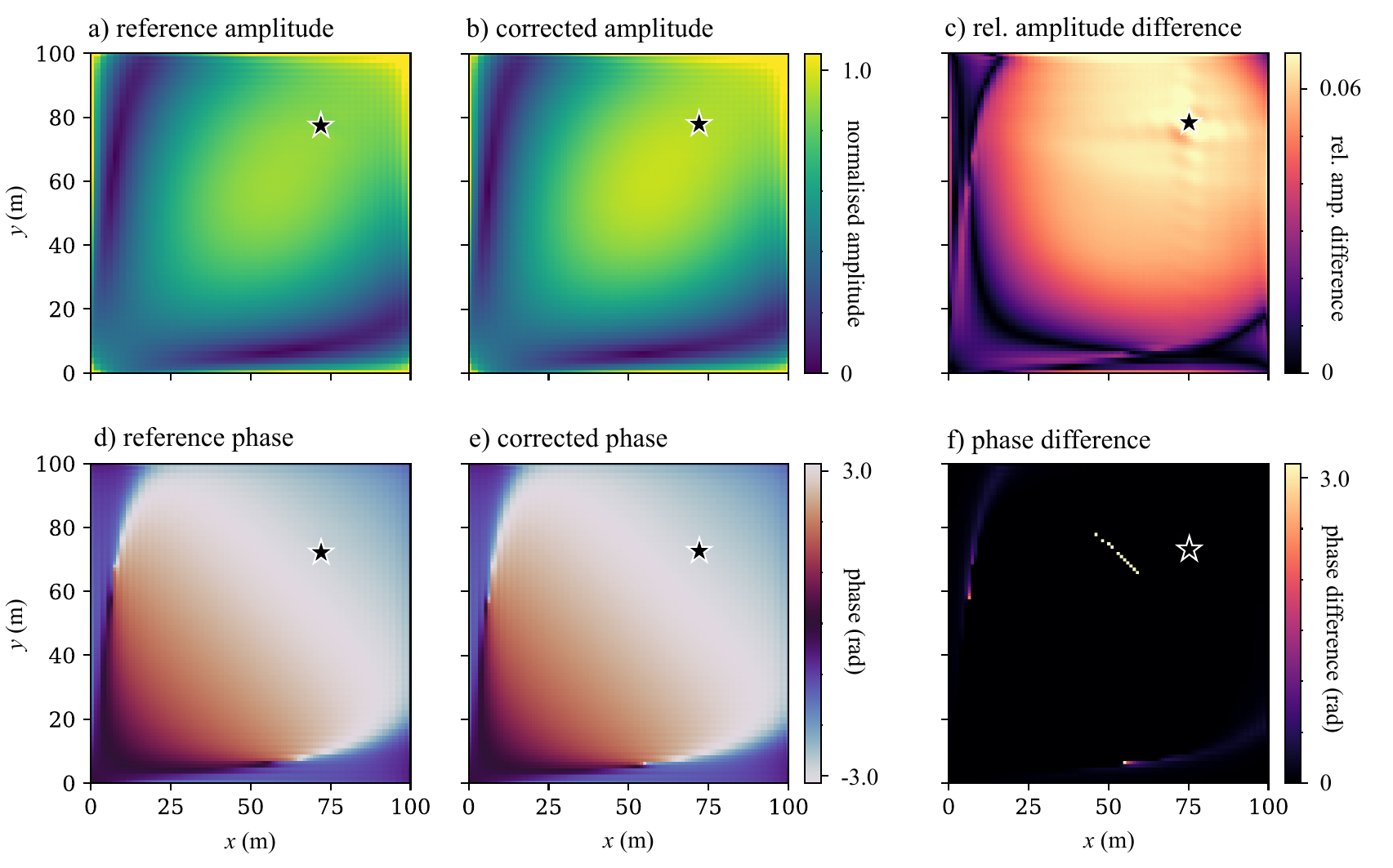}
  \caption{Comparison of amplitude and phase for a vertical-component wavefield snapshot at the surface at $t=0.6\,\mathrm{s}$. The source location is marked by a black star. a-c) Normalised amplitudes of the high-accuracy reference $u_h$, the corrected low-accuracy simulation $\mathcal{N}u_l$, and the difference between the two. Except for the region around the source, amplitude errors are on the order of few percent. d-e) The same as in panels a-c but for the phase (masked in regions of low amplitude).
}
  \label{fig:phase_amp_corr}
\end{figure}

Snapshots of simulated and corrected wavefields at three different times are shown in Fig.~\ref{fig:wavefields}. The low-resolution input $u_l$ exhibits strong numerical dispersion relative to the high-resolution reference $u_h$. In addition to a phase mismatch, the frequency content of $u_l$ is visibly too low, and the amplitudes are too small by a factor of $\sim$2. After correction, the upscaled wavefield $\mathcal{N}u_l$ closely matches the high-resolution simulation $u_h$. Slight differences are visible only in the direct difference $\mathcal{N}u_l - u_h$.\\[5pt]
Fig.~\ref{fig:seismogramLAHA} provides a trace-by-tace comparison of the vertical component wavefield. The nearest receiver is placed 10 m from the source and subsequent receivers spaced at 1 m intervals. The corrected traces align almost perfectly with the high-resolution reference, whereas the low-resolution input shows noticeable delays and amplitude distortions. These results confirm that the neural corrector successfully corrects $u_l$ into a close approximation $\mathcal{N}u_l \approx u_h$.
%
%
%
We conclude this section with a more detailed quantification of the approximation error, presented in Fig.~\ref{fig:phase_amp_corr}. For this, we consider a decomposition into amplitude and phase of vertical-component wavefields at the surface for a snapshot at $t=0.6~\mathrm{s}$. The amplitude panels demonstrate that the neural corrector $\mathcal{N}u_l$ reproduces the high-accuracy reference $u_h$ with residuals typically one order of magnitude smaller than the signal itself. The largest discrepancies occur near the source, where the combined FFT + DCT compression used for dimensionality reduction suppresses high wave number content. In contrast, the phase panels show almost perfect agreement between the neural correction and the reference, with errors close to zero across most of the domain and only minor localized mismatches at later times. These results highlight that the network restores the phase information with a remarkable accuracy.
%

\section{Limitations}\label{S:Limitations}

\begin{figure}
  \centering
  \includegraphics[width=0.8\columnwidth]{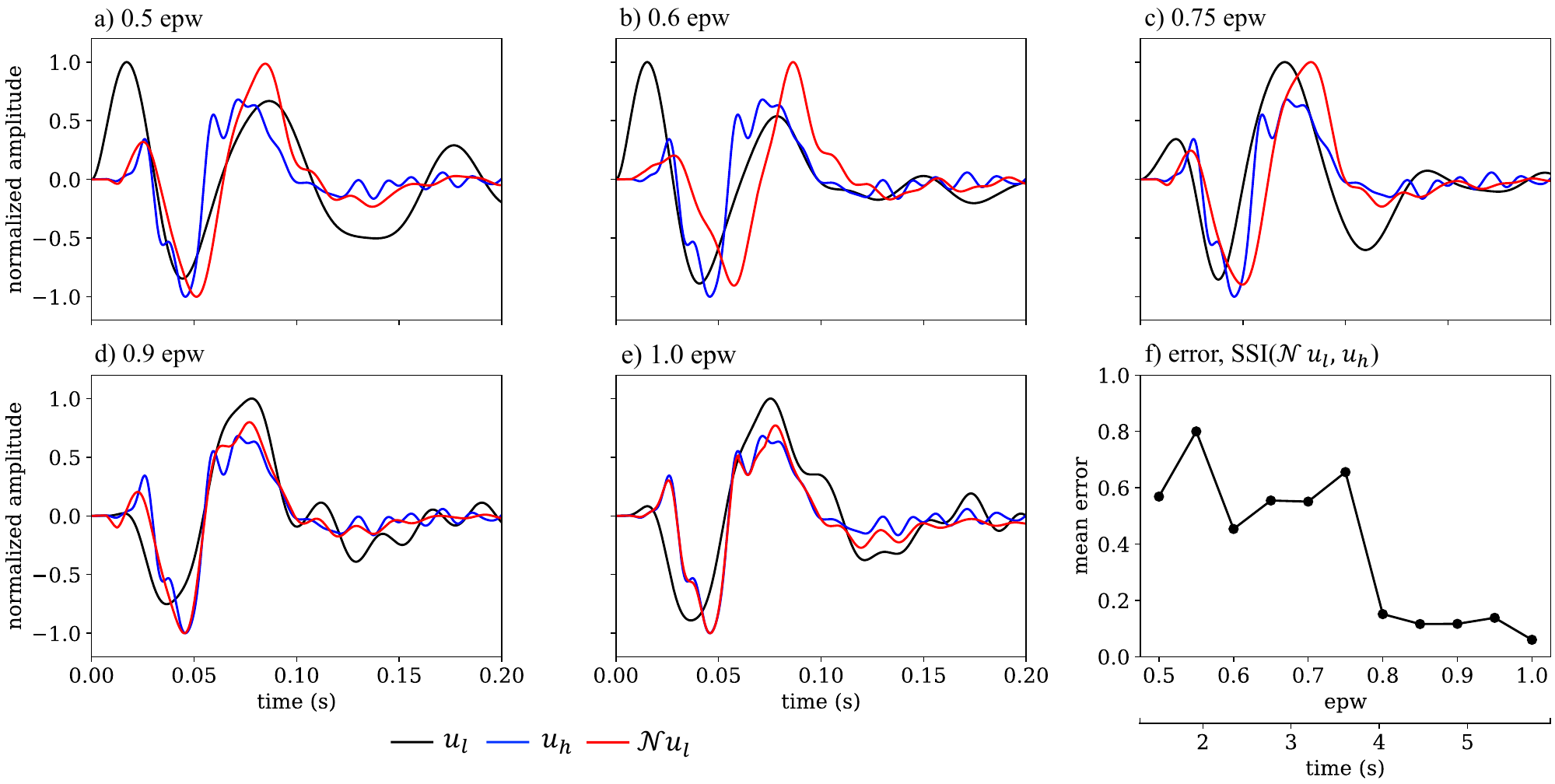}
    \caption{Trace-by-trace comparison of approximation accuracy for input wavefields computed with a number of elements per wavelength that is lower than in the training dataset. a–e) Vertical-component waveforms from the low-accuracy simulation ($u_l$, black), the high-accuracy reference ($u_h$, blue) and neural correction ($\mathcal{N}u_l$, red). The number of elements per wavelength (epw) of the input simulations ranges between 0.5 and 1.0. f) Mean error of the corrected wavefields with respect to the high-accuracy reference simulation as a function of elements per wavelength and corresponding simulation time.}
  \label{fig:acc_lim}
\end{figure}

Despite the promising results of the neural corrector, some limitations must be acknowledged to clarify its range of applicability and to highlight directions for future research. First, the performance of the method depends on user-defined hyperparameters, such as the number of retained Fourier modes, the network depth and width, and the learning rate schedule. We found most of these by trial-and-error. Suboptimal choices may result in over-smoothing, instability, or overfitting.\\[5pt]
Second, the ability of the corrector to generalise beyond the training dataset is naturally limited. We performed training using input simulation data with one element per wavelength. Hence, trying to make corrections with input data computed with less elements per wavelength, is expected to result in lower approximation accuracy. Fig. \ref{fig:acc_lim} quantifies this loss of accuracy when the number of elements per wavelength is smaller than what we used for training. Down to 0.8 elements per wavelength, the neural corrector produces approximations that are similar to those obtained with one element per wavelength, indicating some ability to generalise beyond the training dataset. However, below 0.8 elements per wavelength, the approximation accuracy drops rapidly towards practically useless values.\\[5pt]
Finally, there are limitations related to computational cost and memory. While the combined FFT+DCT representation reduces the memory footprint and accelerates training, it comes at a price: High-wavenumber information is truncated, sharp interfaces are smoothed, and small-scale heterogeneities may be lost. Extending the approach to account for broader frequency ranges or finer discretisations would require increasing the spectral cutoff, which significantly raises memory requirements and training time. We further discuss this issue in section \ref{S:Accuracy}.
%

\section{Discussion}\label{S:Discussion}

The numerical experiments in section~\ref{S:Results} demonstrate that the proposed neural dispersion corrector can substantially reduce the computational cost of wavefield simulations while maintaining high accuracy. To better appreciate its potential and limitations, it is helpful to place this approach within the broader context of surrogate modelling and numerical error mitigation. In the following, we discuss how the method compares to existing strategies in terms of computational efficiency, achievable accuracy, and its relation to other forms of dispersion correction. This perspective highlights both the advantages and the trade-offs that define the applicability of neural dispersion correction.

\subsection{Balancing training and simulation cost}

Our approach complements methods that fully replace the wave equation with a surrogate network by offering a different balance between training and simulation cost. This can be illustrated with a comparison to the work of \citet{Zou_2024} who trained an NO in a setting very similar to ours, i.e., using the Salvus spectral-element solver for the isotropic elastic wave equation to propagate $\sim$10 wavelengths through the computational domain.\\[5pt]
\citet{Zou_2024} report a forward simulation speed-up of $\sim$100, relative to the reference spectral-element simulation. On the order of 30$\,$000 training examples were needed to optimise the network and achieve satisfactory generalisation. Our neural corrector, in contrast, achieves a forward simulation speed-up of 16 in the example from section \ref{S:Results}. The number of training samples was on the order of 1$\,$000.\\[5pt]
This illustrates that the neural dispersion corrector can be interpreted as a mechanism to transfer computational cost between the training and simulation stages. From the perspective of the grid-based reference solution, it reduces simulation cost at the expense of a higher training cost. Conversely, from the perspective of a full network surrogate, it reduces training cost at the expense of increased simulation cost. In summary, this suggests that the niche of neural dispersion correctors may be defined by applications that are limited by memory and storage requirements.

\subsection{Accuracy}\label{S:Accuracy}

The required numerical accuracy is determined by the noise level in the observations to which the simulations are supposed to be compared. With a few tuning parameters, the neural dispersion corrector can be optimised to achieve this accuracy without expending resources on superfluous refinement. In addition to the tuning parameters shared by all neural networks -- depth of the network, number of training epochs and size of the training dataset -- the neural corrector may use frequency subsampling and DCT compression to steer accuracy.\\[5pt]
Frequency subsampling acts to limit the bandwidth over which the network generalises. This should ideally be adjusted to the bandwidth and temporal sampling of the observations. Consequently, an important avenue for improvement concerns the systematic selection of Fourier modes in the FNO layers. At present, we retain a fixed number of modes, but a more principled approach would be to choose the cutoff $K$ per axis as
\begin{equation}\label{E:fno013}
K \;\approx\; \Big\lceil \tfrac{k_{\text{cut}} L}{2\pi} \Big\rceil \quad \text{(FFT)}, 
\qquad 
K \;\approx\; \Big\lceil \tfrac{k_{\text{cut}} L}{\pi} \Big\rceil \quad \text{(DCT)} ,
\end{equation}
with a small safety margin. Here, $L$ is the physical domain, and $k_{\text{cut}} = 2\pi f_{\max}/v_{\min}$ links the maximum frequency and the minimum velocity in the model to the highest spatial wavenumber that must be represented. Incorporating this criterion into network design would better align the retained modes with the spectral content of the training data. Relatedly, frequency subsampling directly controls the bandwidth over which the network generalises. In future work, the subsampling strategy could be explicitly adapted to the bandwidth and temporal sampling of the observations, ensuring that the learned corrector focuses on the physically relevant portion of the spectrum while avoiding unnecessary computational overhead.\\[5pt]
The use of the DCT is more intricate. It acts as a low-pass in space and has the effect of removing sharp features and the associated scattered wavefield components. How the DCT is employed in practice depends on the use case. Owing to the diffraction limit, medium properties are often unknown at lengths scales smaller than about half the minimum wavelength. This sets the DCT wavenumber cutoff and dictates the size of a training example. In other applications, such as non-destructive testing of engineering structures, sharp contrasts may be well-known and required in the simulation. Consequently, the DCT wavenumber cutoff may need to be shifted towards higher wavenumbers, thereby resulting in larger memory requirements and longer training times.

\subsection{Relation to temporal dispersion correctors}

Our method has conceptual similarities with the work of \citet{Wang_2015} and \citet{Koene_2018}, who showed that dispersion errors induced by time discretisation can be removed by analytical filter operations. Their approach permits an increase of $\Delta t$ up to the CFL limit for a fixed spatial discretisation. Hence, computational cost scales as $1/\Delta t$. In addition to temporal discretisation errors, neural correctors also compensate for spatial discretisation errors. While the corrector is not analytic and requires training, the computational cost reduction scales as $1/\Delta t^{d+1} \propto 1/\Delta x^{d+1}$, assuming that $\Delta t$ is increased up to the CFL limit for an increased $\Delta x$.

\section{Conclusions}\label{S:Conclusions}

We proposed a neural operator that corrects the dispersion error of low-accuracy wave simulations, thereby enabling the recovery of high-accuracy results at almost the same computational cost of coarse simulations. In our examples, we consider 3-D elastic (vectorial) wave propagation, and merely used 1$\,$000 training examples and one GPU for network training. Our trained network shows strong generalisation capabilities for arbitrary 3-D media and source locations not seen during the training phase. The neural corrector achieves excellent accuracy in both phase and amplitude for almost the entire testing set.\\[5pt]
Our approach differs from surrogate neural networks that aim to replace the wave equation entirely. Instead of learning the underlying physics, the proposed corrector specifically targets numerical dispersion errors. A further distinguishing aspect is the requirement of only a relatively small dataset to effectively learn the correction operator. To construct a memory-tractable dataset, we incorporate FFT and DCT operations to reduce the dataset dimensionality, leading to substantial savings in memory usage, RAM requirements, and training computational node hours.


\section*{Acknowledgements}
We express our gratitude to Nicola Bienati from ENI for his valuable assistance and insightful comments, which greatly enhanced the quality of this work. Furthermore, we gratefully acknowledge Johan Robertsson for inspiration and Patrick Marty for help with the spectral-element simulations.

\bibliographystyle{unsrtnat}
\bibliography{references}
\end{document}